\pdfoutput=1

\documentclass[11pt,twoside,a4paper,cmspaper,final,collab]{cms-tdr}
\def\svnVersion{182559}\def\cmsCernNo{2012-376}\def\cmsCernDate{\today}\def\cmsMessage{Submitted to Physics Letters B}

\begin{document}\cmsNoteHeader{SMP-12-024}

\hyphenation{had-ron-i-za-tion}
\hyphenation{cal-or-i-me-ter}
\hyphenation{de-vices}

\RCS$Revision: 175699 $
\RCS$HeadURL: svn+ssh://svn.cern.ch/reps/tdr2/papers/SMP-12-024/trunk/SMP-12-024.tex $
\RCS$Id: SMP-12-024.tex 175699 2013-03-10 14:21:53Z alverson $
\newlength\cmsFigWidth
\ifthenelse{\boolean{cms@external}}{\setlength\cmsFigWidth{0.85\columnwidth}}{\setlength\cmsFigWidth{0.4\textwidth}}
\ifthenelse{\boolean{cms@external}}{\providecommand{\cmsLeft}{top}}{\providecommand{\cmsLeft}{left}}
\ifthenelse{\boolean{cms@external}}{\providecommand{\cmsRight}{bottom}}{\providecommand{\cmsRight}{right}}
\newcommand{\tauh}{\ensuremath{\Pgt_\mathrm{h}}}
\newcommand{\taumu}{\ensuremath{\Pgt_{\Pgm}}}
\newcommand{\taue}{\ensuremath{\Pgt_{\Pe}}}
\newcommand{\taus}{\ensuremath{2\ell 2\Pgt}}
\newcommand{\PZ}{\cPZ\xspace}
\newcommand{\ZZ}{\cPZ\cPZ\xspace}
\newcommand{\WW}{\PWp\PWm\xspace}
\newcommand{\WZ}{\PW\cPZ\xspace}
\newcommand{\pp}{\Pp\Pp\xspace}
\newcommand{\Wjets}{\ensuremath{\PW+\text{jets}}\xspace}
\hyphenation{ATGCs}
\newcommand{\fourl}{\ensuremath{\ell^{+}_\mathrm{i} \ell^{-}_\mathrm{i} \ell^{+}_\mathrm{j} \ell^{-}_\mathrm{j}}}
\newcommand{\ZZfourl}{\ensuremath{{\PZ\PZ} \rightarrow 2\ell 2\ell}}
\newcommand{\ZZtwoltwotau}{\ensuremath{{\PZ\PZ} \rightarrow 2l2\tau}}
\providecommand{\mH}{\ensuremath{m_{\PH}\xspace}}
\providecommand{\vecEtm}{\ensuremath{\vec{E}_\mathrm{T}^{\mathrm{miss}}}\xspace}
\newcommand{\dytt}{\ensuremath{\cPZ/\Pgg^*\to\tau^+\tau^-}}
\newcommand{\dyll}{\ensuremath{\cPZ/\Pgg^*\to \ell^+\ell^-}}
\newcommand{\dyee}{\ensuremath{\cPZ/\Pgg^*\to\Pep\Pem}}
\newcommand{\dymm}{\ensuremath{\cPZ/\Pgg^*\to\Pgmp\Pgmm}}
\newcommand{\tw}{\cPqt\PW}
\newcommand{\wgamma}{\PW\Pgg}
\renewcommand{\lumi}{\ensuremath{\,\mathrm{(lum.)}}\xspace}
\cmsNoteHeader{SMP-12-024} 
\title{Measurement of the $\PWp\PWm$ and $\cPZ\cPZ$ production cross sections
in pp collisions at $\sqrt{s} = 8\TeV$ }

\date{\today}

\abstract{
The  $\PWp\PWm$ and $\cPZ\cPZ$ production cross sections are measured
in proton-proton collisions at $\sqrt{s} = 8\TeV$ with the CMS experiment
at the LHC in data samples corresponding to an integrated luminosity of up to 5.3\fbinv.
The measurements are performed in the leptonic
decay modes $\PWp\PWm \to \ell'\nu\ell''\nu$ and $\cPZ\cPZ \to 2\ell 2\ell'$,
where $\ell = \Pe, \Pgm$ and $\ell'(\ell'') = \Pe, \Pgm, \Pgt$.
The measured cross sections
$\sigma ( \pp \to \PWp\PWm) =  69.9 \pm 2.8 \stat \pm 5.6 \syst \pm 3.1 \lumi\unit{pb}$
and $\sigma ( \pp \to \cPZ\cPZ) =  8.4 \pm 1.0 \stat \pm 0.7 \syst \pm 0.4 \lumi\unit{pb}$,
for both $\cPZ$ bosons produced in the mass region $60 < m_{\cPZ} < 120\GeV$,
are consistent with standard model predictions.
These are the first measurements of the diboson production  cross sections
at $\sqrt{s} = 8\TeV$.
}

\hypersetup{%
pdfauthor={CMS Collaboration},%
pdftitle=
{Measurement of W+W- and ZZ production cross sections in pp collisions at sqrt(s)=8 TeV},%
pdfsubject={CMS},%
pdfkeywords={CMS, physics, W and Z pair production}}

\maketitle

\section{Introduction}

The study of $\WW$ and $\cPZ\cPZ$ production in proton-proton collisions provides an
important test of the standard model (SM).
Any deviations of the
measured cross sections from SM predictions would indicate new physics.
Measurements of electroweak $\WW$ and $\cPZ\cPZ$ production are essential for an accurate
estimate of irreducible
backgrounds for Higgs boson studies.

Previous measurements of $\WW$ and $\cPZ\cPZ$ production were
performed at the Large Hadron Collider (LHC) at a centre-of-mass
energy $\sqrt{s}=7\TeV$.
With a data set corresponding to an integrated luminosity of $36\unit{pb}^{-1}$,
the Compact Muon Solenoid (CMS) Collaboration
measured the $\WW$ cross section
$\sigma ( \pp \to \WW) = 41.1 \pm 15.3 \stat \pm 5.8 \syst \pm 4.5 \lumi\unit{pb}$~\cite{CMSWWxsTGC35pb},
in good agreement with the SM prediction of $47  \pm 2\unit{pb}$ from Ref.~\cite{MCFM1}.
The ATLAS Collaboration measured
$\sigma ( \pp \to \WW) = 51.9 \pm 2.0 \stat \pm 3.9 \syst \pm 2.0 \lumi\unit{pb}$~\cite{ATLAS:2012mec}
using $4.6\fbinv$ of data.
The $\cPZ\cPZ$ cross section measurement from CMS used $5\fbinv$ of data;
the measured value,
$\sigma ( \pp \to \PZ\PZ) =  6.24 \,^{+0.86}_{-0.80} \stat \,^{+0.41}_{-0.32} \syst \pm 0.14 \lumi\unit{pb}$,
is consistent with the SM prediction of
$6.3 \pm 0.4\unit{pb}$ for both $\cPZ$ bosons in the mass range
$60 < m_{\cPZ} < 120\GeV$~\cite{SMP-12-007}.
ATLAS measured
$\sigma ( \pp \to \cPZ\cPZ) = 6.7 \pm 0.7 \stat \, ^{+0.4}_{-0.3} \syst \pm 0.3 \lumi\unit{pb}$~\cite{ATLASnewXS}
with a data sample corresponding to an integrated luminosity of $4.6\fbinv$.
Measurements of the $\WW$ and $\cPZ\cPZ$ cross sections performed at
the Tevatron are summarized in Refs.~\cite{CDFZZxs,D0ZZxs, CDFWWxs, D0WWxslvjj, D0WWxsllvv}.
All measurements are found to agree well with the corresponding SM predictions.

In this Letter, the first measurements of the $\WW$ and $\PZ\PZ$
production cross sections at $\sqrt{s} = 8\TeV$ are presented.
The analysis is based on
data collected in 2012 with the CMS experiment at the LHC,
corresponding to an
integrated luminosity of $3.5\fbinv$ for the $\WW$ measurement and  $5.3\fbinv$
for the $\PZ\PZ$ measurement.
The measurements are performed in the
$\WW \to \ell'\nu\ell''\nu$  and
$\PZ\PZ \to 2\ell 2\ell'$ decay channels, where
$\ell$ is $\Pe$ or $\Pgm$,
and $\ell'(\ell'')$ is $\Pe$, $\Pgm$, or $\Pgt$.
If a $\Pgt$ lepton is present in the $\WW$ final state,
only leptonic decays of the $\Pgt$ lepton are considered.
If a $\Pgt$ lepton is present in the $\PZ\PZ$ final state, one $\PZ$ is required to decay either
into
$\Pep\Pem$
or $\Pgmp\Pgmm$,
and the second
$\PZ$ into $\Pgt^+\Pgt^-$ in four possible final states: $\tauh \tauh$, $\taue \tauh$, $\taumu \tauh$, and
$\taue \taumu$,
where $\tauh$ indicates a $\Pgt$ lepton decaying
hadronically, and $\taue$ and $\taumu$ indicate taus decaying into
an electron and a muon, respectively.

The SM background sources to the $\WW$ event sample
include
$\PW\gamma^{(*)}$, top-quark
($\ttbar$ and $\tw$), $\dyll$, and diboson
($\PW\PZ$ and $\PZ\PZ$) production, as well as
$\PW$+jets and QCD multijet events, where at
least one of the jets is misidentified as a lepton.
The SM background sources to the $\PZ\PZ$ event sample
include contributions from $\cPZ\cPqb\cPaqb$ and $\ttbar$ processes,
where the final states contain two isolated leptons and two $\cPqb$ jets with secondary leptons,
and from $\cPZ$+jets and $\PZ\PW$+jets processes where the jets are misidentified
as leptons.

\section{The CMS detector and simulation}

While the CMS detector is described in detail elsewhere~\cite{CMSExperiment}, the
key components for this analysis are summarized here.
The CMS experiment uses a right-handed coordinate system, with the
origin at the nominal interaction point, the $x$ axis
pointing to the center of the LHC ring, the $y$ axis pointing
up (perpendicular to the plane of the LHC ring), and the $z$ axis
along the anticlockwise-beam direction. The polar angle
$\theta$ is measured from the positive $z$ axis and the
azimuthal angle $\phi$ is measured in the $x$-$y$ plane.
The magnitude of the transverse
momentum ($\PT$) is calculated as $\PT = \sqrt{p_x^2 + p_y^2}$.
A superconducting solenoid occupies the
central region of the CMS detector, providing an axial magnetic
field of 3.8\unit{T} parallel to the beam direction.
The silicon pixel and strip
tracker, the crystal electromagnetic calorimeter, and the brass/scintillator hadron
calorimeter are located within the solenoid. A quartz-fibre
Cherenkov calorimeter extends the coverage to $|\eta| <$ 5.0, where pseudorapidity
is defined as $\eta=-\ln[\tan{(\theta/2)}]$.
Muons are measured in gas ionization detectors embedded
in the steel flux return yoke outside the solenoid.
The first level of the CMS trigger system, composed of custom
hardware processors, is designed to select the most interesting events
in less than 3\mus using information from the calorimeters and muon
detectors.
The high-level-trigger processor farm decreases the event rate from 100\unit{kHz}
delivered by the first level trigger to a few hundred hertz, before data storage.

Several Monte Carlo (MC) event generators are used to simulate the signals and
backgrounds. The $\WW$ production via $\Pq\Paq$ annihilation is generated
with the \MADGRAPH~\cite{Madgraph5} event generator, and
\PYTHIA~\cite{Sjostrand:2006za} is used for parton showering, hadronization, and
the underlying event simulation.
The $\Pg\Pg \to \WW$ process, which is expected to
contribute 3\% of the total $\WW$ production rate~\cite{MCFM},
is generated with \textsc{gg2ww}~\cite{ggww}.
The $\ZZ$ production via $\Pq\Paq$ annihilation is generated at
next-to-leading order (NLO) with
\POWHEG2.0~\cite{Alioli:2008gx,Nason:2004rx,Frixione:2007vw}.
The $\Pg\Pg \to \ZZ$ process
is simulated
with {\textsc{gg2zz}}~\cite{Binoth:2008pr}.
Other diboson processes ($\PW\cPZ$, $\PW\gamma^{(*)}$)
and the $\cPZ$+jets samples
are generated with \MADGRAPH. The $\ttbar$ and $\tw$ events are generated at NLO with
\POWHEG.
For leading-order (LO) generators, the default set of parton distribution functions
(PDF) used to produce these samples is CTEQ6L~\cite{CTEQ66}, while
CT10~\cite{ct10} is used for NLO generators.
The $\Pgt$ lepton decays are generated with {\TAUOLA}~\cite{Jadach:1993hs}.
For all processes, the detector response is simulated using a detailed
description of the CMS detector, based on the \GEANTfour~
package~\cite{GEANT}, and event reconstruction is performed with
the same algorithms as used for data.
The simulated samples include additional interactions per bunch crossing (pileup).
The simulated events are weighted so that the pileup distribution matches the data,
with an average pileup of about 20 interactions per bunch crossing.
\section{Event reconstruction}

Both the $\WW$ and $\cPZ\cPZ$ event selections begin with
the reconstruction and identification of lepton candidates.
Electrons are reconstructed
by combining information from the electromagnetic calorimeter
 and tracker~\cite{Baffioni:2006cd,CMS-PAS-EGM-10-004}.
Their identification relies on a multivariate
technique that combines observables sensitive to the amount of bremsstrahlung along the
electron trajectory, the geometrical and momentum matching between the electron trajectory
in the tracker and the energy deposit in the calorimeter,
as well as the shower shape~\cite{Baffioni:2006cd}.
Muons are reconstructed~\cite{CMS-PAS-MUO-10-002}
with information from both the tracker and the muon spectrometer,
and are required to pass selection criteria  similar to those described in Ref.~\cite{CMSWWxsTGC35pb}.
A particle-flow (PF) technique~\cite{CMS-PAS-PFT-09-001} is used
to reconstruct $\tauh$ candidates
with the ``hadron plus strip'' (HPS) algorithm~\cite{Chatrchyan:2011xq}, which
is designed to optimize the performance of
$\tauh$ identification and reconstruction
by considering specific $\tauh$ decay modes.
In the
PF approach, information from all subdetectors is combined
to reconstruct and identify particles produced in the collision.
The particles are classified into mutually exclusive
categories:
charged hadrons, photons, neutral hadrons, muons, and
electrons.
These particles are used to reconstruct
the $\tauh$ candidates; the
neutrinos produced in all $\Pgt$ decays escape detection
and are ignored in the $\tauh$ reconstruction.
The lepton candidates are required to be consistent with the primary vertex of the
event, which is chosen as the vertex with highest $\sum \pt^2$ of its associated tracks.
This criterion provides the correct assignment for the
primary vertex in more than 99\% of both signal and
background events for the pileup distribution observed in the data.

Charged leptons from $\PW$ and $\cPZ$ boson decays are usually isolated
from other activity in the event. For each electron or muon candidate, a cone
is constructed around the track direction at the event vertex.  The scalar
sum of the transverse momenta of all reconstructed
particles consistent with the
chosen primary vertex and contained within the cone is calculated,
excluding the contribution from the lepton candidate itself.
To improve the discrimination
against nonisolated muons from the $\PW$+jets background in the $\WW$ selection,
this procedure is repeated with several cones of different widths.
This isolation information is then
combined by means of a multivariate technique.
For both electrons and muons a correction is applied to account for
the energy contribution
in the isolation cone due to pileup.
A median transverse energy due to pileup
is determined event by event and is
subtracted
from the transverse energy ($\ET$) in the
isolation cone~\cite{Cacciari:subtraction}.
A similar technique is used to form $\Pgt$ lepton isolation quantities.

Jets are reconstructed from the PF particles
using the anti-\kt
clustering algorithm~\cite{antikt} with distance parameter of 0.5,
as implemented in the \textsc{fastjet}
package~\cite{Cacciari:fastjet1,Cacciari:fastjet2}. The jet energy is corrected
for pileup in a manner similar to the correction of the energy inside a lepton
isolation cone. Jet energy corrections are also applied as a function of the jet
$\pt$ and $\eta$~\cite{cmsJEC}.
A multivariate selection is applied to separate jets coming from the
primary interaction from those reconstructed using energy deposits associated with
pileup.
This discrimination is based on the differences in the jet shapes and
in the relative multiplicity of charged and neutral components.
Tracks associated with a jet are required to be consistent with the primary vertex.

To suppress the top-quark background in events without high-\pt jets,
top-quark-tagging techniques
are defined with two methods.
The first method vetoes events containing
muons originating from b quarks~\cite{btag2} appearing in top-quark decays.
The second method uses b-jet tagging applied to jets with
$15 < \pt < 30\GeV$ based on tracks with large impact parameter within
jets.

The missing transverse energy $\vecEtm$ is defined as
the negative vector sum of the
transverse momenta of all reconstructed particles in the event.
A \textit{projected $\MET$} is defined
as (i) the magnitude  of the $\vecEtm$
 component transverse to the closest lepton, if $\Delta \phi(\ell, \vecEtm ) < \pi/2$,
or (ii) the magnitude of the $\vecEtm$ otherwise.
This observable more efficiently rejects $\dytt$ background events in which
the $\vecEtm$
is preferentially aligned with the leptons,
and
$\dyll$ events with mismeasured
$\vecEtm$.
Since the {projected $\MET$} resolution is degraded as pileup increases,
the minimum of two different observables is used: the first includes all particle candidates in the
event, while the second uses only the charged
particle candidates associated with the primary vertex.

\section{Event selection and background estimates}
\label{sec:eventsel_backgrounds}
\subsection{\texorpdfstring{\WW}{WW} production}

Events are selected with two oppositely charged electron or muon
candidates,
both with $\pt > 20\GeV$ and with
$|\eta| < 2.5$ for the electrons
and $|\eta| <  2.4$ for the muons.
The $\tau$ leptons contribute to the measurement
only if they decay to electrons or muons that
pass the selection requirements.
At the trigger level, events are
required to have a pair of electrons or muons where
one of the leptons has $\pt > 17\GeV$ and the other $\pt > 8\GeV$,
or a single electron (muon) with $\pt > 27~(24)\GeV$.
The trigger efficiency is approximately 98\%
for both $\cPq\cPaq\to\WW$ and $\Pg\Pg \to \WW$ processes.

To reduce the background from top-quark decays, events with one or more jets
surviving the jet selection criteria and with $\pt > 30\GeV$ and
$|\eta|< 4.7$ are rejected. The residual top-quark background
is further suppressed by 50\% after applying the top-quark-tagging techniques.

In order to reduce the Drell--Yan background, the {projected $\MET$} is
required to be above 45\GeV in the $\Pep\Pem$ and $\Pgmp\Pgmm$ final states.
For the $\Pe^\pm\Pgm^\mp$ final state, which has a lower
contamination from $\dyll$ decays, the threshold is reduced to $20\GeV$.
These requirements remove more than 99\% of the Drell--Yan background.

To further reduce the Drell--Yan background in the
$\Pep\Pem$ and $\Pgmp\Pgmm$ final states,
the angle in the transverse plane between the dilepton system total momentum and
the most energetic jet with $\pt > 15\GeV$ is required to be smaller than 165 degrees.
Events with dilepton masses
within ${\pm}15\GeV$ of
the \Z mass
or below 12\GeV are also rejected.
Finally, the transverse momentum of the dilepton system, $\pt^{\ell\ell}$,
is required to be above 45\GeV to reduce contributions from Drell--Yan
background  and
events containing jets misidentfied as leptons.

To reduce the background from other diboson processes, such as $\WZ$ or $\ZZ$
production, any event is rejected if it has a third lepton with
$\pt > 10\GeV$ passing the
identification and isolation requirements.
The $\wgamma^{(*)}$ production in which the photon converts is suppressed
by rejecting electrons consistent with a photon conversion.

The $\PW$+jets and QCD multijet backgrounds are estimated from a control region
in which one lepton passes the nominal requirements, while
the other passes looser criteria on impact parameter and isolation, but fails the
nominal requirements. The contribution to the control region
from processes with two genuine leptons is subtracted by using simulation.
The number of events in the signal region is obtained by
scaling the number of events in the control region with the efficiency
for loosely identified lepton candidates to pass the tight selection.
These efficiencies
are measured in data using multijet events and are parametrized according
to the $\pt$ and $\abs{\eta}$ of the lepton candidate.

The normalization of the top-quark background that survives the
top-quark-tagging requirements is estimated from data
by counting the number of top-tagged events and applying the
corresponding top-tagging efficiency.
The top-tagging efficiency
is measured with a control sample dominated by
$\ttbar$ and $\tw$ events, which is selected by requiring a b-tagged jet.

We estimate the Drell--Yan contribution to the $\Pep\Pem$ and
$\Pgmp\Pgmm$ final states outside of the Z mass window by normalising the
event yield from simulation to the observed number of events inside the $\Z$ mass window.
The methods used to estimate both the top-quark and Drell--Yan
backgrounds are described in more detail in Ref.~\cite{Chatrchyan:2012ty}.

Finally, a control sample with three reconstructed leptons is used to measure the
data-to-MC scaling factor
for the $\wgamma^{(*)}$ process.
We use only $\wgamma^{(*)} \rightarrow \ell\nu\Pgmp\Pgmm$ events
for this measurement because the $\ell\nu\Pep\Pem$ final state is difficult to
separate from other backgrounds.
This measurement is used to normalise the simulated $\wgamma^{(*)}$ background
contribution from asymmetric gamma decays in which one lepton escapes
detection~\cite{wgammastart}.

Other backgrounds, such as $\WZ$ and $\ZZ$ diboson production, are estimated
from simulation.

The  $\PW$+jets and $\wgamma^{(*)}$ background
estimate is checked using data events that pass all the selection
requirements with the exception that the two leptons must have the same charge.
After subtraction of the expected
$\WZ$ background,
this sample
is dominated by $\PW$+jets and $\wgamma^{(*)}$ events.
The  $\dytt$ contamination is
checked using $\dyee$ and $\dymm$ events selected in data, where the
leptons are replaced with simulated $\Pgt$ lepton decays.

\subsection{\texorpdfstring{\ZZ}{ZZ} production}

Selected events are required to have at least one electron or muon
with
$\pt > 20\GeV$ and another one with $\pt > 10\GeV$, and
$|\eta| < 2.5~(2.4)$ for electrons (muons).
All other electrons (muons) are required to have $\pt > 7~(5)\GeV$.
The $\tauh$ candidates are required to have $\pt > 20\GeV$ and $|\eta| <  2.3$.
All leptons
must originate from the same vertex
and be isolated.
At the trigger level, events are
required to have a pair of electrons or muons,
one lepton with $\pt > 17\GeV$ and the other with $\pt > 8\GeV$.

The selected events are required to contain two $\PZ$ candidates.
One candidate, denoted by $\PZ_1$, should decay into
electrons or muons,
$\PZ \to \ell^+\ell^-$, and must have
reconstructed  invariant mass
$ 60 < m_{\ell\ell} < 120\GeV$.
If more than one candidate is found, the one with mass closest
to the $\PZ$ mass is considered as $\PZ_1$.

The selection requirements for the second $\cPZ$ candidate,
denoted by $\cPZ_2$, depend on the final state.
In the $4\Pgm$, $4\Pe$, and $2\Pe 2\Pgm$
final states the isolation requirements are the same as for the leptons
from $\PZ_1$.
For the $\Pep\Pem\taue^{\pm}\taumu^{\mp}$ and
$\Pgmp\Pgmm\taue^{\pm}\taumu^{\mp}$ final states
the electron and muon $\pt$ values are required
to exceed 10\GeV.
In final states with $\cPZ_2 \to \taue\tauh, \taumu\tauh$
the isolation requirements
for all the electrons and
muons are tighter.
A study of inclusive $\cPZ \to \Pgt^+\Pgt^-$ production~\cite{CMS-EWK-TAU}
shows
that modifying the electron and muon
isolation requirements is  a more effective way to reduce background in such final
states than
imposing tighter isolation criteria on $\tauh$.

The invariant mass of the reconstructed $\cPZ_2$
is required to satisfy
$ 60 < m_{\ell^+\ell^-} < 120\GeV$ when $\cPZ_2$ decays into $\Pep\Pem$ or $\Pgmp\Pgmm$.
In the $\taus$ final states,
the visible invariant mass of the
reconstructed $\PZ_2 \to \Pgt^+\Pgt^-$ is shifted to smaller
values because of undetected neutrinos in $\tau$ decays.
Therefore,
in the final states involving $\tau$ leptons, the visible mass
 is required to satisfy
$ 30 < m_{\Pgt^+\Pgt^-} < 90\GeV$, and
the leptons from the same $\PZ$ are required to be
separated by $\Delta R > 0.4$ for the $\PZ_1$, and
by $\Delta R > 0.5$ for the $\PZ_2$.

Estimated acceptances
of the selection requirements
defined
with respect to the full phase space
are 58\%, 56\%, 54\%, and 25\%
for the $4\Pe$, $2\Pe 2\Pgm$, $4\Pgm$, and $2\ell2\Pgt$ final states, respectively.

The major contributions to the background come from
$\cPZ$ production in association with jets,
$\PW\cPZ$ production in association with
jets, and $\ttbar$.
In all these cases, a
jet or nonisolated lepton is misidentified as an isolated
electron, muon, or $\tauh$.
The relative contribution of each source of background
depends on the final state.

For the background estimation, two different approaches are used.
Both start by relaxing the isolation and identification criteria for
two additional reconstructed
lepton objects indicated as $\ell_\text{reco}\ell_\text{reco}$ in the $\PZ_1 + \ell_\text{reco}\ell_\text{reco}$
event sample.
The additional pair of leptons is required to have like-sign
charge (to avoid signal contamination)
and same flavour ($\Pe^{\pm}\Pe^{\pm}, \Pgm^{\pm}\Pgm^{\pm}$).
The first method estimates the number of \cPZ+X background events in the signal region by taking
into account the lepton misidentification probability for each of the two additional leptons.
The second method uses a control region with two
opposite-sign leptons that fail the isolation
and identification criteria. The background in the signal region
is estimated
by weighting the events in the control region
with the lepton misidentification probability.
In addition, a control region with  three passing leptons and one failing lepton is used to
account for contributions from backgrounds with three prompt leptons and one misidentified lepton.
Comparable background rates in the signal region are found within the uncertainties from
both methods.

\section{Systematic uncertainties}

The uncertainty in the signal acceptance for the two measurements
due to variations in the parton distribution functions and the
value of $\alpha_{s}$ is estimated by following
the PDF4LHC
prescription~\cite{Botje:2011sn}.
Using CT10~\cite{ct10}, MSTW08~\cite{Martin:2009iq}, and NNPDF~\cite{nnpdf} sets,
the uncertainties are $2.3\%~(4.0\%)$
for the $\Pq\Paq \to \WW(\cPZ\cPZ)$ processes.

The effects of higher-order corrections are found
by varying the QCD renormalisation
 and factorisation  scales
simultaneously up and down by a factor of two
using the \MCFM program~\cite{MCFM}.
The variations in the acceptance are found to be
1.5\% for the $\Pq\Paq \to \WW$ process and to be
negligible for the $\Pq\Paq \to \PZ\PZ$ and
$\Pg\Pg \to \PZ\PZ$ processes.

The $\WW$ jet veto efficiencies in data are estimated from simulation, and multiplied by a
data-to-simulation scale factor derived from $\dyll$ events in the $\cPZ$ peak:
$\epsilon_{\WW}^\text{data} = \epsilon_{\WW}^\mathrm{MC} \times \epsilon_{\cPZ}^\text{data}/\epsilon_{\Z}^\mathrm{MC}.$
The uncertainty is thus factorized into the uncertainty in the $\cPZ$ efficiency in data and the
uncertainty in the ratio of the $\WW$ efficiency to the \cPZ\ efficiency in simulation
($\epsilon_{\WW}^\mathrm{MC}/\epsilon_{\Z}^\mathrm{MC}$). The former, which is statistically dominated,
is 0.3\%. Theoretical uncertainties due to higher-order corrections contribute most to the
$\WW/\cPZ$ efficiency ratio uncertainty, which is estimated to be 4.6\% for $\WW$ production. The
data-to-simulation correction factor is found to be close to one and is not applied.

Simulated events are scaled according to the lepton
efficiency correction factors
measured using data control samples, which are typically close to one.
The uncertainties in the measured
identification and isolation efficiencies are
found to be 1--2\% for muons and electrons, and 6--7\%
for $\tauh$. The uncertainty in the trigger efficiency is 1.5\%.
The uncertainty in the lepton energy scale is about 3\% for $\tauh$, and 1--2.5\%
and 1.5\% for electrons and muons, respectively.

The uncertainties in the $\PZ$+jets, $\PW\PZ$+jets, and $\ttbar$ backgrounds
to the $\PZ\PZ \to 2\ell2\ell'$ selection are 30--50\%
depending on the decay channel. These uncertainties comprise the
statistical and systematic uncertainties in the misidentication
rates measured in data control samples.

The systematic uncertainties in the $\PW$+jets, $\cPZ$+jets, and top backgrounds to
the $\WW \to \ell'\nu\ell''\nu$
selection are 36\%, 24\%, and 15\%, respectively.
The theoretical uncertainties in the $\WZ$ and $\ZZ$ cross sections are calculated
following the same prescription as for the signal acceptance.
Including the experimental uncertainties gives a systematic uncertainty in
$\WZ$ and $ZZ$ backgrounds  of approximately 10\%.

The uncertainty assigned to the pileup reweighting procedure
amounts to $2.3\%$.
The uncertainty in the integrated luminosity is 4.4\%~\cite{lumiPAS12}.
\section{Results}
\subsection{\texorpdfstring{\WW}{WW} cross section measurement}

The observed and expected signal plus background yield is summarized in
Table~\ref{tab:wwselection_all}. The expected
$\WW$ contribution is calculated assuming the
SM cross section.

\begin{table}[tbh]
  \begin{center}
    \topcaption{Expected and observed event yields for the $\WW$ selection.
    The
    uncertainties correspond to the statistical and systematic uncertainties added
    in quadrature.
    \label{tab:wwselection_all}}
   \begin{tabular}{lc}
        \hline
        Channel                  & $\ell'\nu\ell''\nu$ \\ \hline
        $\WW$               &  $684  \pm  50$    \\ \hline
        \ttbar and t\PW       &  $132  \pm 23$    \\
        $\PW$+jets          &  $60  \pm  22$    \\
        $\PW\cPZ$ and $\cPZ\cPZ$      &   $27  \pm 3$    \\
        $\cPZ/\gamma^*$+jets         &   $43  \pm 12$    \\
	$\PW\gamma^{(*)}$ &   $14  \pm  5$    \\
        Total background        &  $275  \pm 35$  \\ \hline
        Signal + background     &  $959  \pm 60$  \\ \hline
        Data                    & 1111                \\ \hline
      \end{tabular}
  \end{center}
\end{table}

The total background yield is $275 \pm 35$
events and the expected signal and background yield is
$959 \pm 60$ events, with 1111 events observed.

The measured $\WW$ yield is calculated by subtracting the
estimated contributions of the various
 background processes.  The product of the
 signal efficiency and acceptance averaged over all
lepton flavors including $\tau$ leptons is $(3.2 \pm 0.2)\%$.
Using the W $\to \ell \nu$ branching ratio of $0.1080 \pm 0.0009$
from Ref.~\cite{pdg}, the $\WW$ production cross section in pp collision data at
$\sqrt{s} = 8\TeV$ is measured to be
\ifthenelse{\boolean{cms@external}}{
\begin{multline*}
\sigma ( \pp \to \PWp\PWm) =\\
69.9 \pm 2.8 \stat \pm 5.6 \syst \pm 3.1 \lumi\unit{pb}.
\end{multline*}
}
{
\begin{equation*}
\sigma ( \pp \to \PWp\PWm) =
69.9 \pm 2.8 \stat \pm 5.6 \syst \pm 3.1 \lumi\unit{pb}.
\end{equation*}
}
The statistical uncertainty reflects the total number of observed events.
The systematic uncertainty includes both the statistical and systematic uncertainties in the background prediction,
as well as the uncertainty in the signal efficiency.
This measurement is slightly higher than the
SM expectation of
$57.3\, ^{+2.3}_{-1.6}\unit{pb}$, calculated in Ref.~\cite{MCFM1}
by using {\MCFM\xspace} at NLO with the MSTW08 PDF and setting the factorization
and renormalization scales to the $\PW$ mass.
Additional processes may increase the production yield in the $\WW$ final state by
as much as 5\% for the event selection used in this analysis.
Higgs boson production would give an
additional contribution of about
4\% of the cross section given above, based on
next-to-next-to-leading-order cross section calculations for the
$\PH\to\WW$ process~\cite{LHCHiggsCrossSectionWorkingGroup:2011ti} under
the assumption that the newly discovered resonance~\cite{cmshig:2012gu,atlashig:2012gk}
is a SM-like Higgs boson with a mass of 125\GeV.
Contributions from diffractive production~\cite{Bruni:1993is},  double
parton scattering, and QED exclusive
production~\cite{Pukhov:2004ca} are also considered.

\begin{figure*}[hbtp]
\begin{center}
{\includegraphics[width=0.45\textwidth]{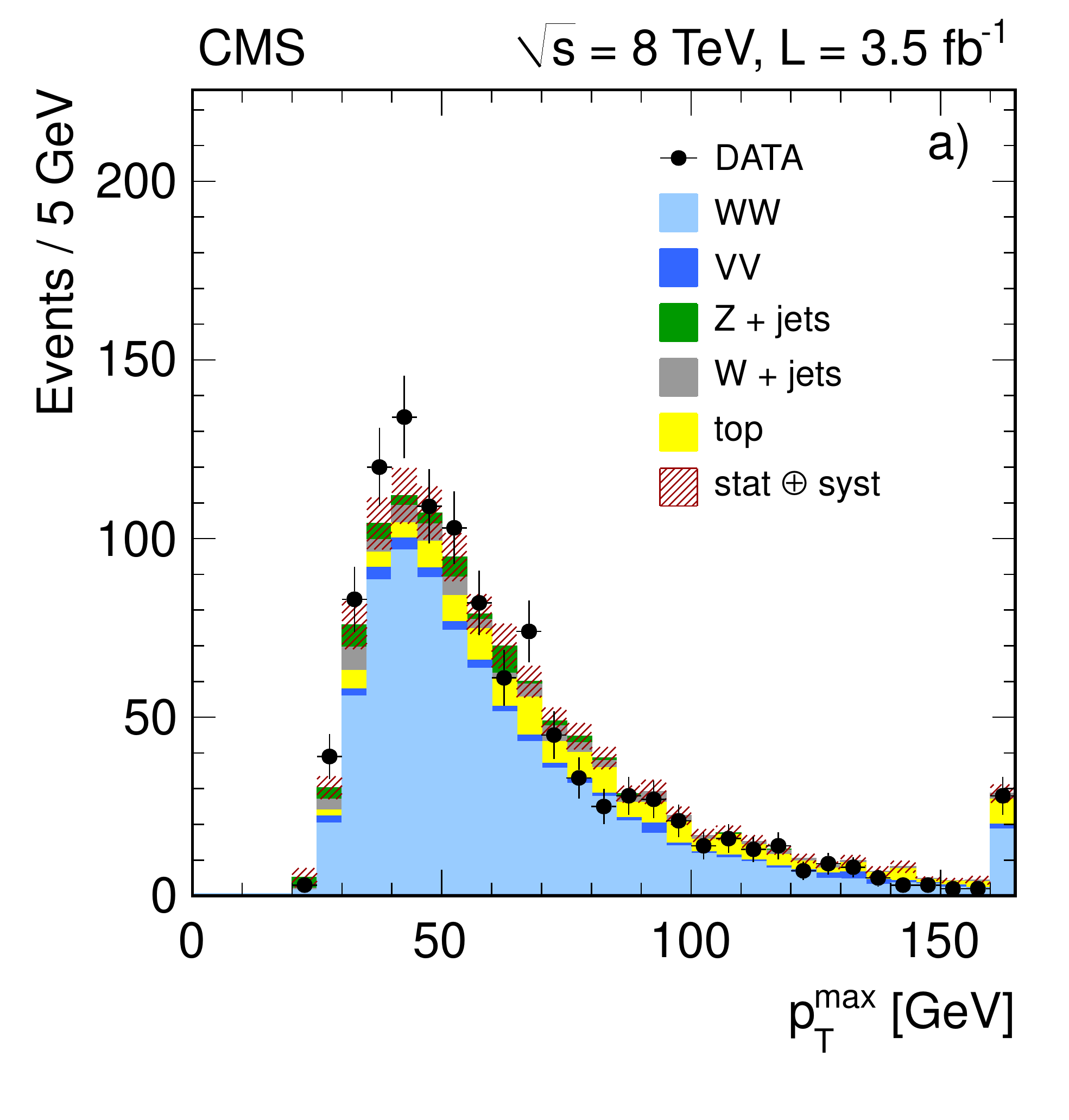}}
{\includegraphics[width=0.45\textwidth]{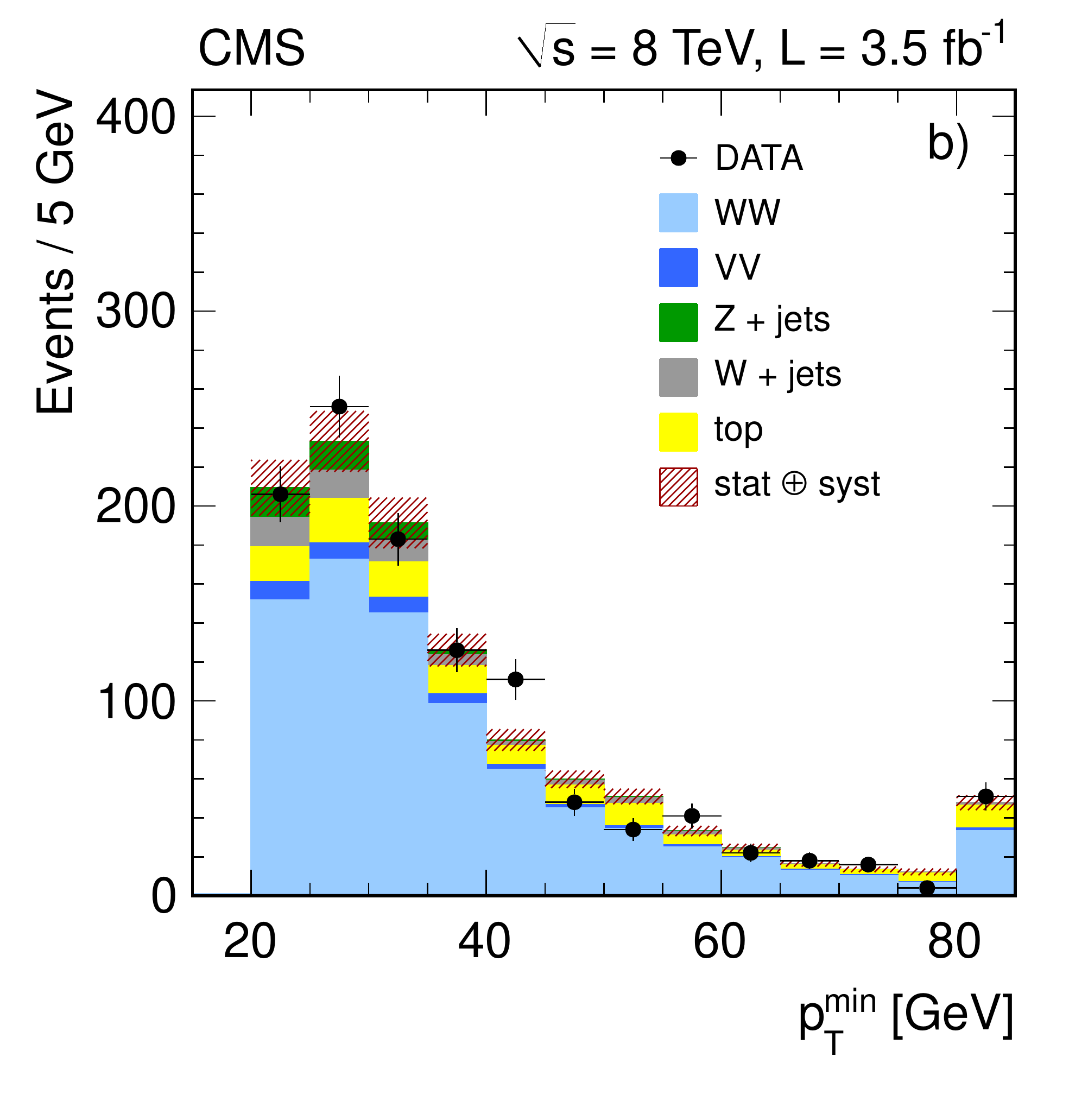}}
{\includegraphics[width=0.45\textwidth]{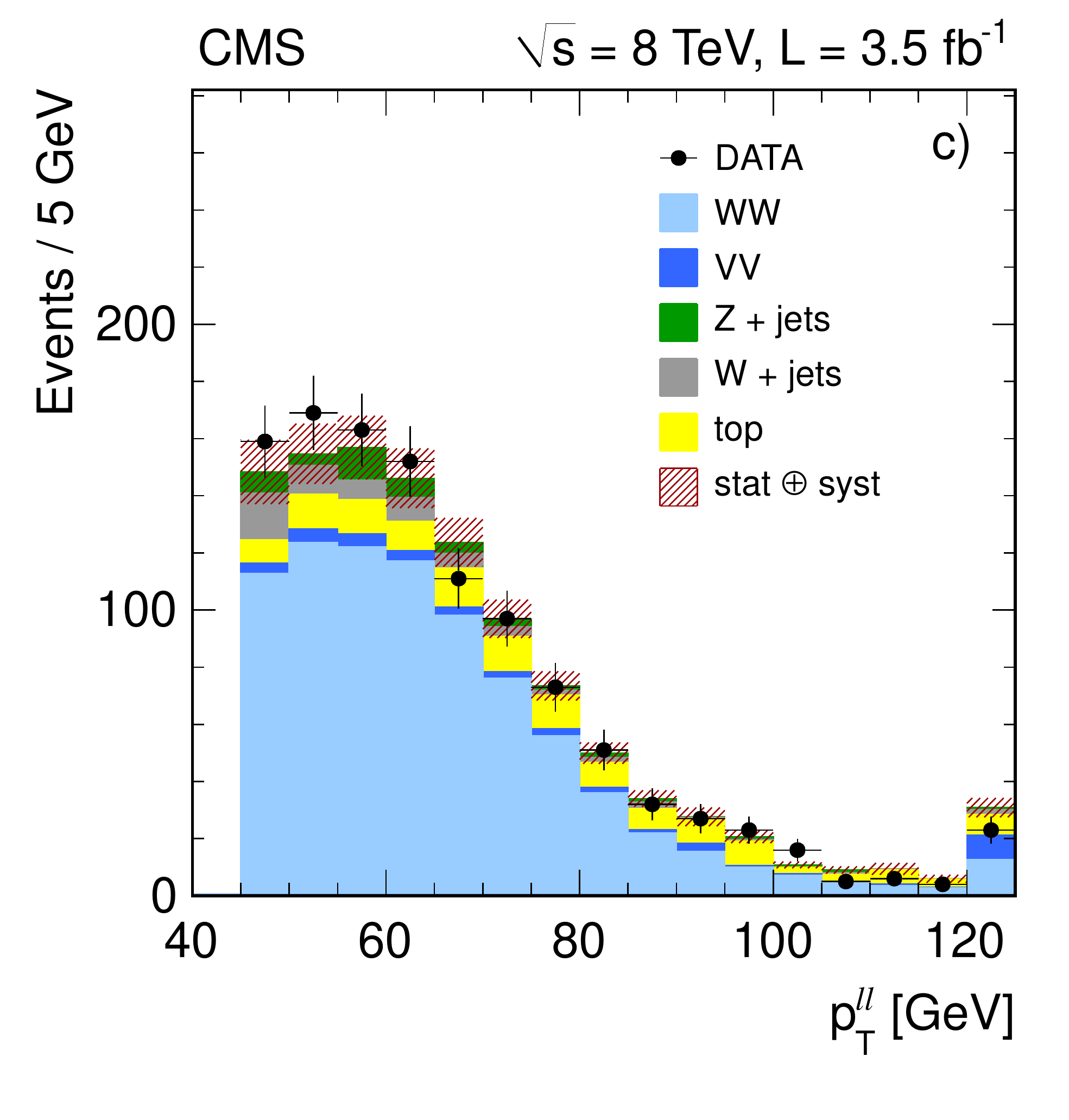}}
{\includegraphics[width=0.45\textwidth]{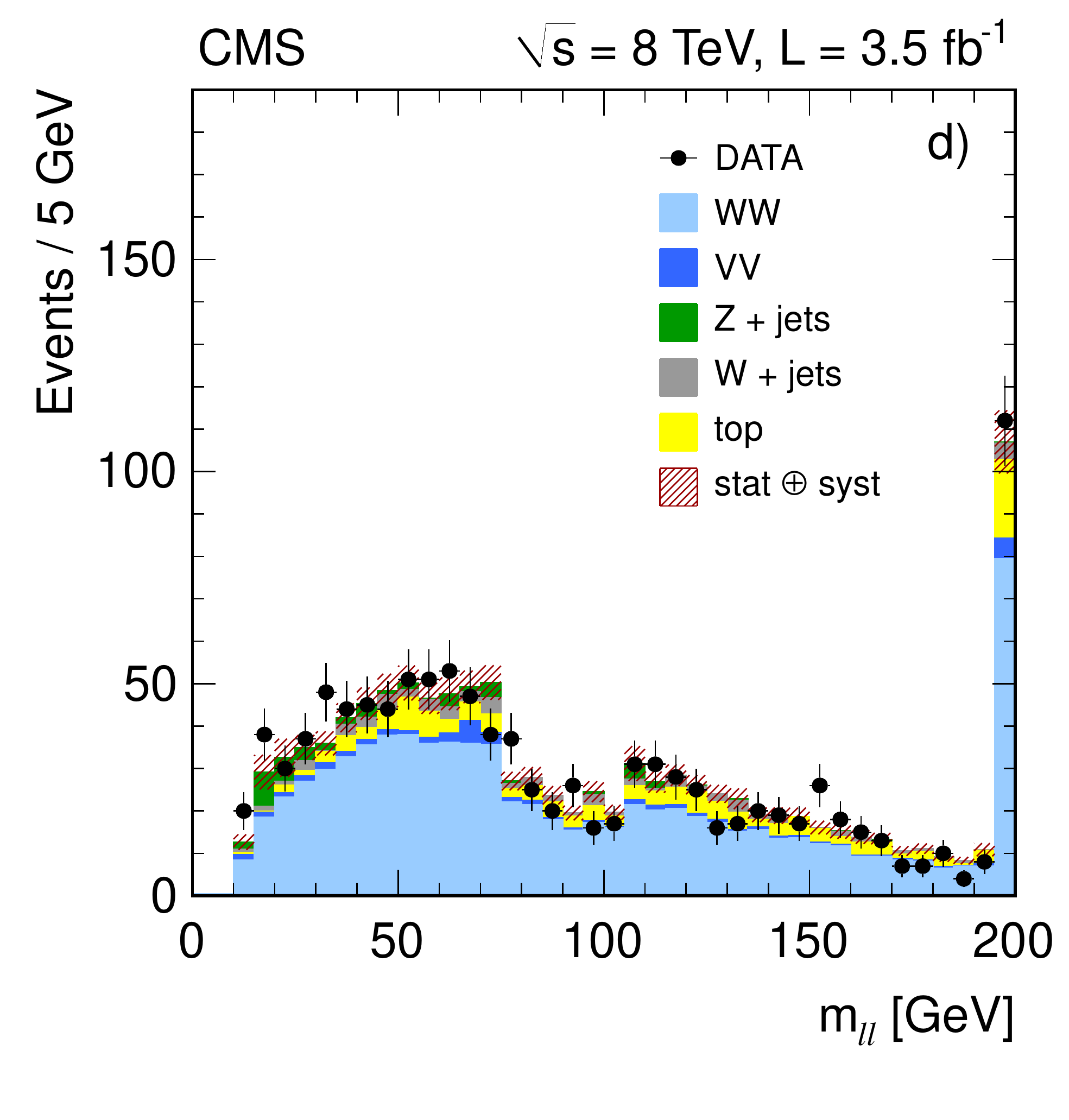}}
\caption{Distributions for $\WW$ candidate events of (a) the leading lepton transverse momentum $\pt^\text{max}$,
(b) the trailing lepton transverse momentum $\pt^\text{min}$,
(c) the dilepton transverse momentum $\pt^{\ell\ell}$, and (d) the dilepton invariant mass $m_{\ell\ell}$.
Points represent the data, and shaded histograms represent
the $\WW$ signal and the background processes. The
last bin includes the overflow.
The $\WW$ signal is scaled to the measured cross section,
and the background processes are normalized to
the corresponding estimated values in Table~\ref{tab:wwselection_all}.
 \label{fig:distributions}}
 \end{center}
\end{figure*}

The distributions of the leading lepton transverse momentum $\pt^\text{max}$,
the trailing lepton transverse momentum $\pt^\text{min}$,
the dilepton transverse momentum $\pt^{\ell\ell}$, and the dilepton invariant mass $m_{\ell\ell}$
are shown in
Fig.~\ref{fig:distributions},
where the $\WW$ contribution
is normalized to the measured cross section.

\subsection{\texorpdfstring{\ZZ}{ZZ} cross section measurement}

\begin{figure*}[htbp]
\begin{center}
{\includegraphics[width=0.45\textwidth]{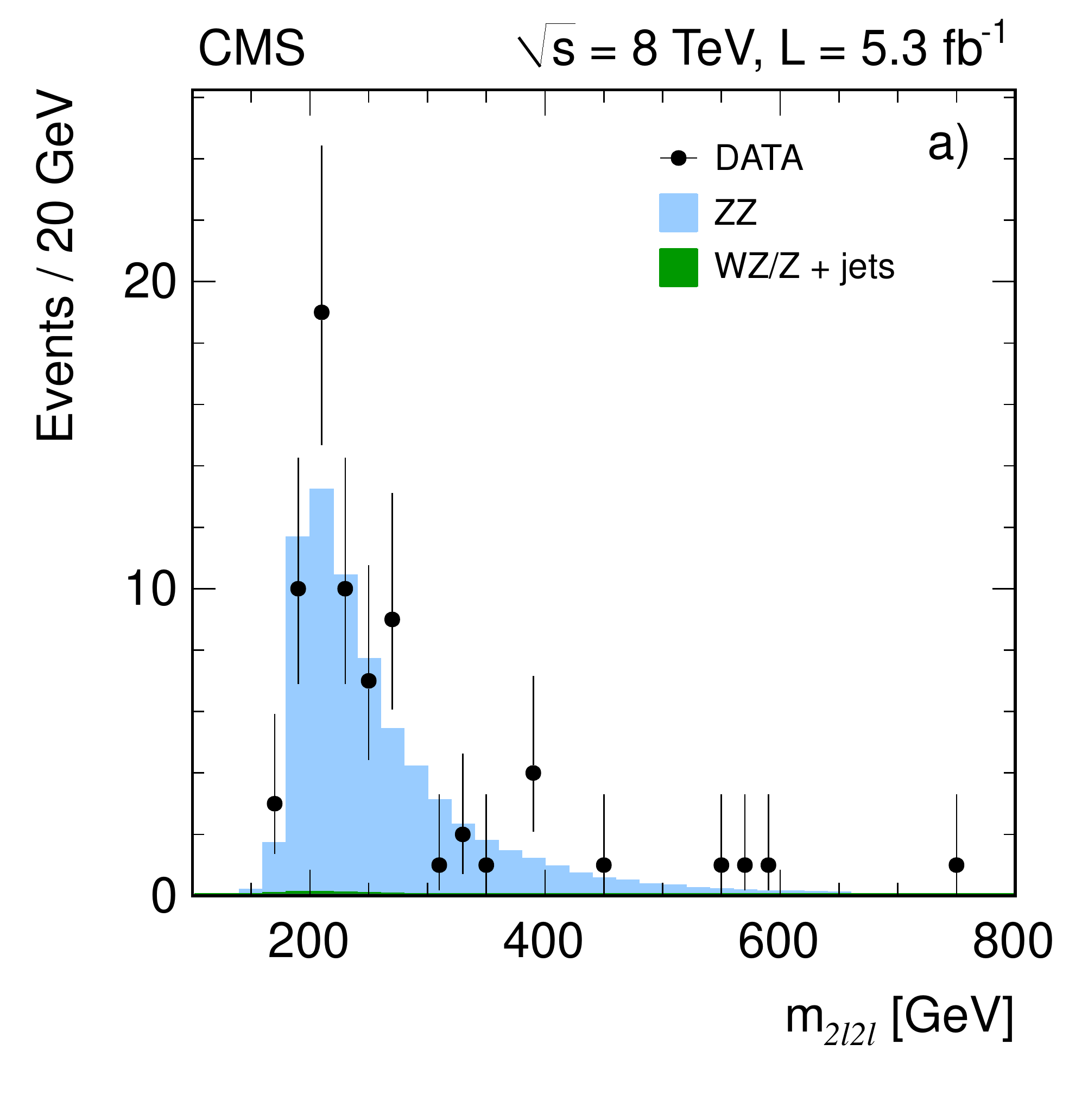}}
{\includegraphics[width=0.45\textwidth]{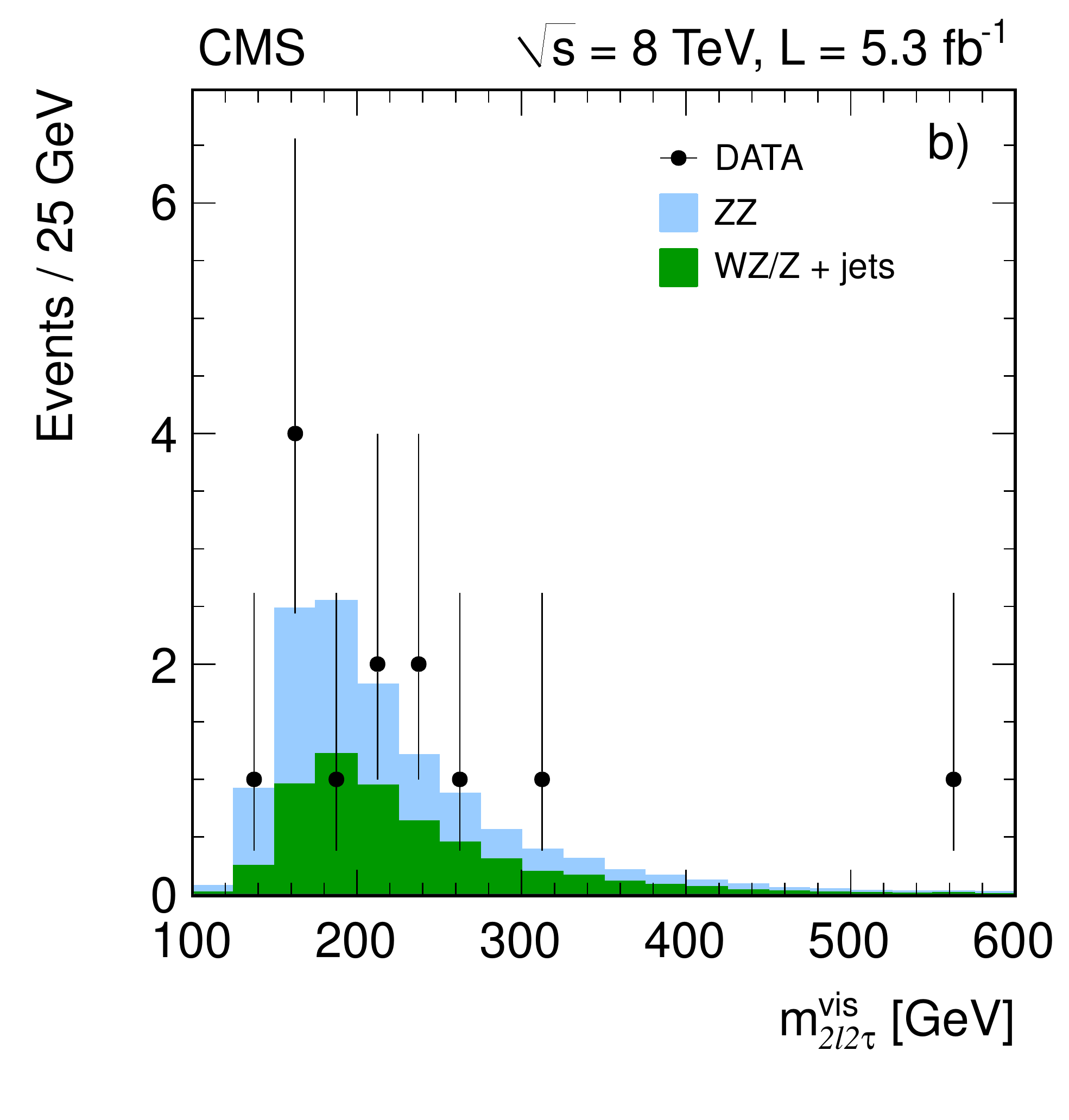}}
{\includegraphics[width=0.45\textwidth]{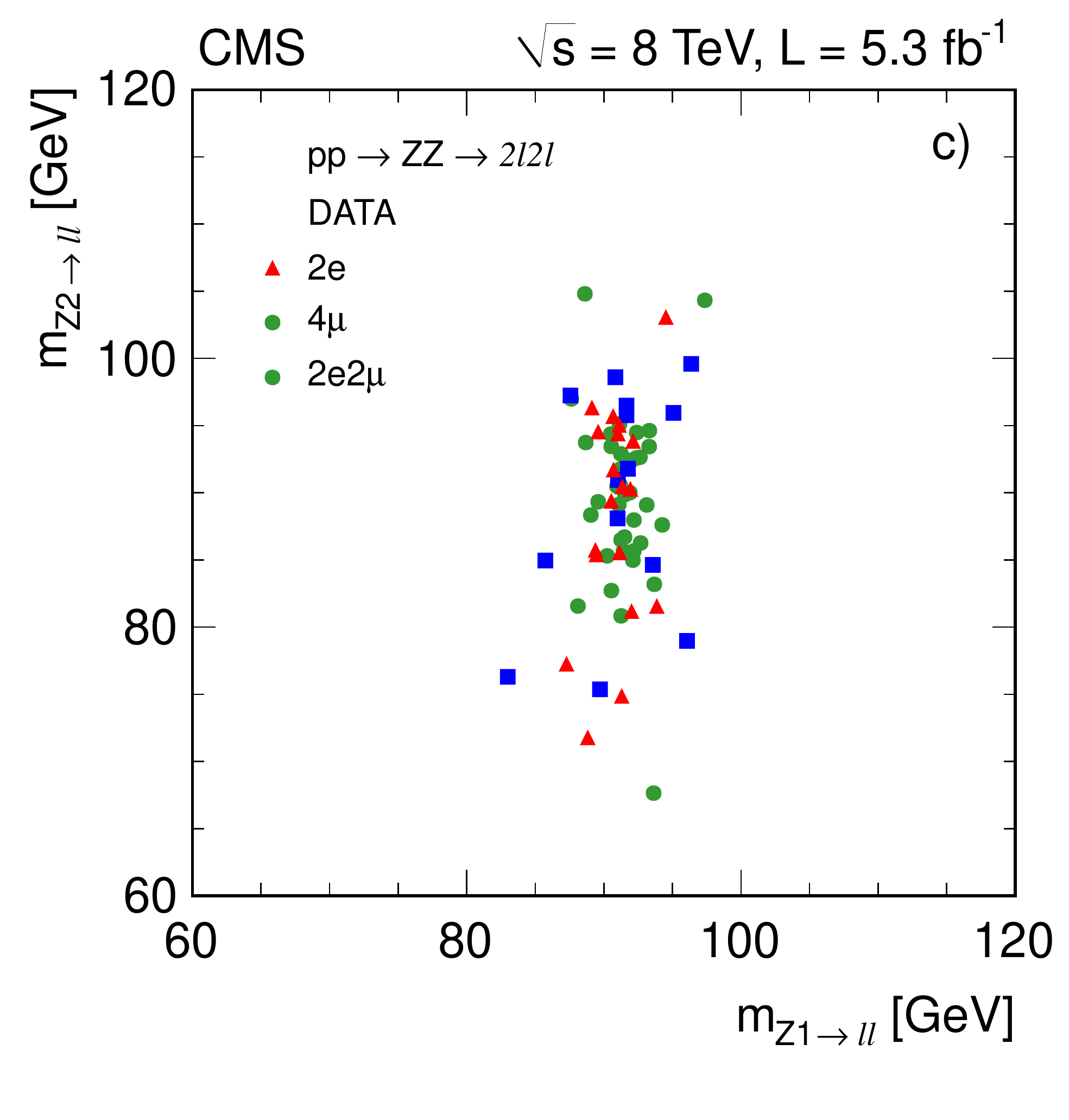}}
{\includegraphics[width=0.45\textwidth]{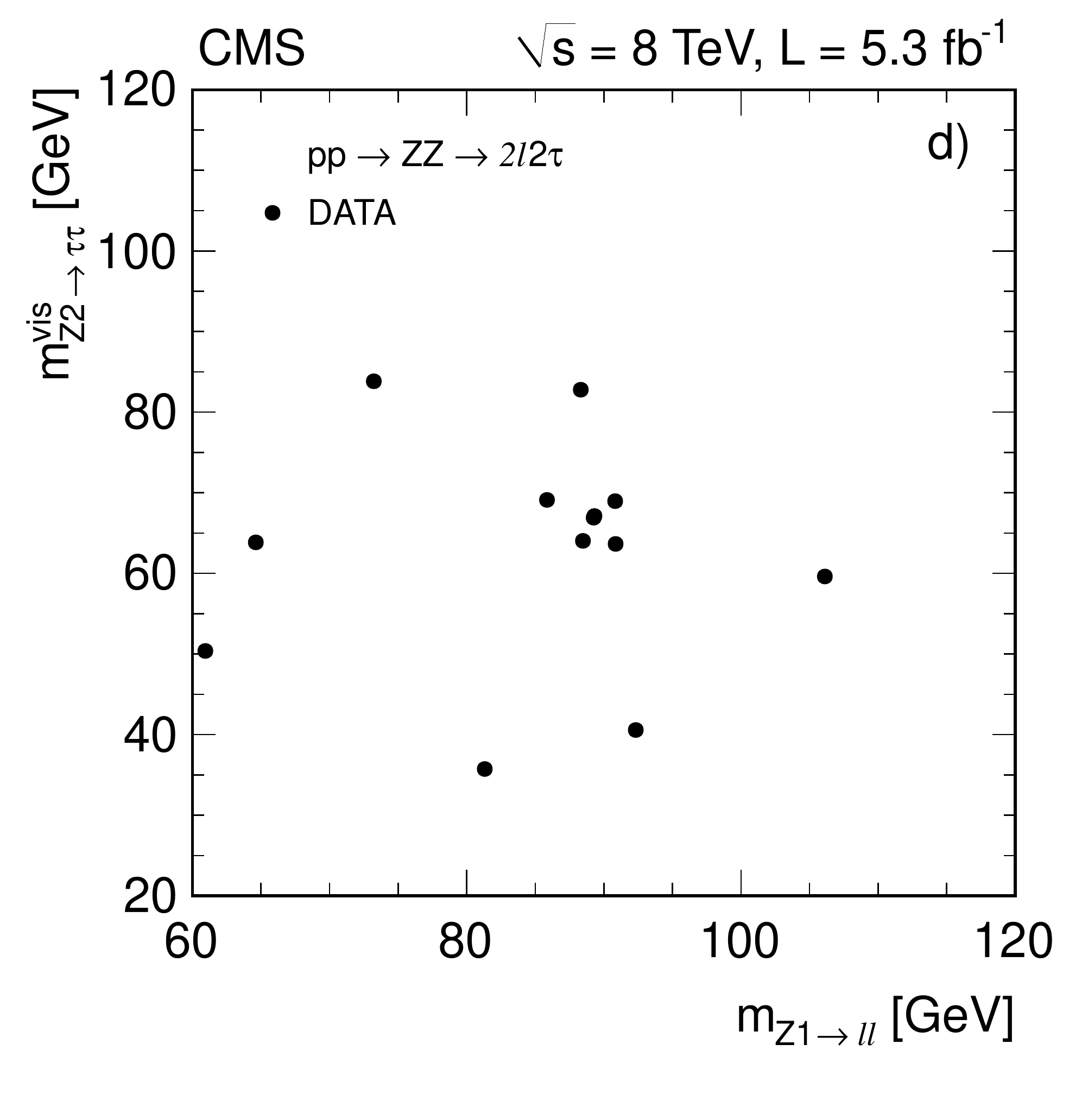}}
\caption{Distributions for $\ZZ$ candidate
 events of (a) the four-lepton reconstructed mass for the sum of the
$4\Pe, 4\Pgm$, and $2\Pe2\Pgm$ channels and (b) the sum of the $\taus$ channels.
Points represent the data, and shaded histograms represent
the expected $\PZ\PZ$ signal and the reducible background.
The shapes of the signal and background are taken from the MC simulation, with each
component normalized to
the corresponding estimated value from Table~\ref{table:results}.
The distributions (c) and (d) demonstrate the relationship between
the reconstructed $\PZ_1$ and $\PZ_2$ masses.
Different symbols are used to present different decay channels.}
\label{fig:Mass4l}
\end{center}
\end{figure*}

Table~\ref{table:results} presents the observed
and expected yields and the number of the estimated background events
in the signal region.
There are 71 candidates observed
in the
$4\Pe, 4\Pgm$, and $2\Pe2\Pgm$ channels, to be
compared to an expectation of $65.6 \pm 4.4 $ events.
Among the expected events
1.4 are from background
processes.
In the $\taus$ channels 13 candidates are observed.
The expected $12.1 \pm 1.6$ events for $\taus$ channels contain
5.6 events from background processes.
The reconstructed four-lepton invariant mass distributions
are shown in Figs.~\ref{fig:Mass4l}(a) and (b) for the sum of the
$4\Pe, 4\Pgm$, and $2\Pe2\Pgm$ channels, and the sum of
all the $\taus$ channels.
Data are compared to the SM expectations.
The shapes of the signal and the background are taken from the MC simulation, with each
component normalized to
the corresponding estimated value from Table~\ref{table:results}.
The reconstructed masses in $\taus$ states are shifted downwards
with respect to the generated values by about $30\%$ because of the
undetected  neutrinos in $\tau$ decays.
Figures~\ref{fig:Mass4l}(c) and (d) demonstrate the relationship between
reconstructed $\PZ_1$ and $\PZ_2$ masses.

\begin{table*}[htb]
\begin{center}
\topcaption{
Expected and observed event yields for the $\ZZ$ selection.
The
uncertainties correspond to the statistical and systematic uncertainties added
in quadrature.
}
\label{table:results}
\begin{tabular}{lcccc}
\hline
Channel & $4\Pe$ & $4\Pgm$ & $2\Pe2\Pgm$ & $2\ell2\Pgt$  \\
\hline
\cPZ\cPZ & 11.6 $\pm$ 1.4   &  20.3 $\pm$ 2.2 & 32.4 $\pm$ 3.5  &  6.5 $\pm$ 0.8  \\ 
Background  & 0.4 $\pm$ 0.2   &  0.4 $\pm$ 0.3 & 0.5 $\pm$ 0.4  &  5.6 $\pm$ 1.4   \\ 
\hline
Signal+background  & 12.0 $\pm$ 1.4 & 20.7 $\pm$ 2.2 & 32.9 $\pm$ 3.5  & 12.1 $\pm$ 1.6  \\ 
\hline
Data & 14 & 19  & 38 & 13 \\ 
\hline
\end{tabular}
\end{center}
\end{table*}

To measure the $\PZ\PZ$ cross section the numbers of observed events are unfolded
in a combined likelihood fit.
Each decay mode is treated as a separate channel giving
eleven measurements to combine: $4\Pe, 4\Pgm$, $2\Pe2\Pgm$, and eight $\taus$ channels.
The $\Pgt$-lepton decay modes are treated separately;
the methodology used for event reconstruction and selection
ensures that the decay modes are mutually exclusive.
The joint likelihood is a combination of the likelihoods for the
individual channels, which include
the signal and background hypotheses.
The statistical and systematic uncertainties are introduced in the form of
nuisance parameters
via log-normal distributions around the estimated central values.
The resulting cross section is
\ifthenelse{\boolean{cms@external}}{
\begin{multline*}
\sigma ( \pp \to \ZZ) = \\ 8.4 \pm 1.0 \stat \pm 0.7 \syst \pm 0.4 \lumi\unit{pb}.
\end{multline*}
}
{
\begin{equation*}
\sigma ( \pp \to \ZZ) =  8.4 \pm 1.0 \stat \pm 0.7 \syst \pm 0.4 \lumi\unit{pb}.
\end{equation*}
}
This is
to be compared to the theoretical value of
$7.7 \pm 0.4\unit{pb}$
calculated with \MCFM
at NLO for $ \Pq\Paq \to \cPZ\cPZ$ and
LO for $\Pg\Pg \to \cPZ\cPZ$
with
MSTW08 PDF, and factorization and renormalization scales set to the $\PZ$ mass, for both lepton pairs in the mass range $60 < m_{\PZ} < 120\GeV$.

\section{Summary}

The $\WW$ and $\cPZ\cPZ$ production
cross sections have been measured
in proton-proton collisions at $\sqrt{s} = 8\TeV$ in the $\WW \to \ell'\nu\ell''\nu$ and
$\cPZ \cPZ \to 2\ell 2\ell'$ decay modes
with $\ell = \Pe, \Pgm$ and $\ell'(\ell'') = \Pe, \Pgm, \Pgt$.
The data samples correspond to an integrated luminosity
of 3.5\fbinv for the $\WW$ and 5.3\fbinv
for the $\PZ\PZ$ measurements.
The measured production cross sections
$\sigma ( \Pp\Pp \to \WW) =  69.9 \pm 2.8 \stat \pm 5.6 \syst \pm 3.1 \lumi\unit{pb}$
and $\sigma ( \Pp\Pp \to \cPZ\cPZ) =  8.4 \pm 1.0 \stat \pm 0.7 \syst \pm 0.4 \lumi\unit{pb}$,
for both $\PZ$ bosons produced in the mass region $60 < m_{\PZ} < 120\GeV$,
are consistent with the standard model predictions.
This is the first measurement of the diboson production cross sections at
 $\sqrt{s} = 8\TeV$.

\section*{Acknowledgements}

We congratulate our colleagues in the CERN accelerator departments for the excellent performance of the LHC and thank the technical and administrative staffs at CERN and at other CMS institutes for their contributions to the success of the CMS effort. In addition, we gratefully acknowledge the computing centres and personnel of the Worldwide LHC Computing Grid for delivering so effectively the computing infrastructure essential to our analyses. Finally, we acknowledge the enduring support for the construction and operation of the LHC and the CMS detector provided by the following funding agencies: BMWF and FWF (Austria); FNRS and FWO (Belgium); CNPq, CAPES, FAPERJ, and FAPESP (Brazil); MEYS (Bulgaria); CERN; CAS, MoST, and NSFC (China); COLCIENCIAS (Colombia); MSES (Croatia); RPF (Cyprus); MoER, SF0690030s09 and ERDF (Estonia); Academy of Finland, MEC, and HIP (Finland); CEA and CNRS/IN2P3 (France); BMBF, DFG, and HGF (Germany); GSRT (Greece); OTKA and NKTH (Hungary); DAE and DST (India); IPM (Iran); SFI (Ireland); INFN (Italy); NRF and WCU (Republic of Korea); LAS (Lithuania); CINVESTAV, CONACYT, SEP, and UASLP-FAI (Mexico); MSI (New Zealand); PAEC (Pakistan); MSHE and NSC (Poland); FCT (Portugal); JINR (Armenia, Belarus, Georgia, Ukraine, Uzbekistan); MON, RosAtom, RAS and RFBR (Russia); MSTD (Serbia); SEIDI and CPAN (Spain); Swiss Funding Agencies (Switzerland); NSC (Taipei); ThEPCenter, IPST and NSTDA (Thailand); TUBITAK and TAEK (Turkey); NASU (Ukraine); STFC (United Kingdom); DOE and NSF (USA).

\vspace*{2ex} 
\bibliography{auto_generated}   

\cleardoublepage \appendix\section{The CMS Collaboration \label{app:collab}}\begin{sloppypar}\hyphenpenalty=5000\widowpenalty=500\clubpenalty=5000\textbf{Yerevan Physics Institute,  Yerevan,  Armenia}\\*[0pt]
S.~Chatrchyan, V.~Khachatryan, A.M.~Sirunyan, A.~Tumasyan
\vskip\cmsinstskip
\textbf{Institut f\"{u}r Hochenergiephysik der OeAW,  Wien,  Austria}\\*[0pt]
W.~Adam, E.~Aguilo, T.~Bergauer, M.~Dragicevic, J.~Er\"{o}, C.~Fabjan\cmsAuthorMark{1}, M.~Friedl, R.~Fr\"{u}hwirth\cmsAuthorMark{1}, V.M.~Ghete, N.~H\"{o}rmann, J.~Hrubec, M.~Jeitler\cmsAuthorMark{1}, W.~Kiesenhofer, V.~Kn\"{u}nz, M.~Krammer\cmsAuthorMark{1}, I.~Kr\"{a}tschmer, D.~Liko, I.~Mikulec, M.~Pernicka$^{\textrm{\dag}}$, D.~Rabady\cmsAuthorMark{2}, B.~Rahbaran, C.~Rohringer, H.~Rohringer, R.~Sch\"{o}fbeck, J.~Strauss, A.~Taurok, W.~Waltenberger, C.-E.~Wulz\cmsAuthorMark{1}
% ...
\vskip\cmsinstskip
\dag:~Deceased\\
1:~~Also at Vienna University of Technology, Vienna, Austria\\
2:~~Also at CERN, European Organization for Nuclear Research, Geneva, Switzerland\\
% ...

\end{sloppypar}
\end{document}